\documentclass[a4paper,english,12pt]{article}

\usepackage[version=3]{mhchem} 
\usepackage{fontenc}           
\usepackage{hyperref}
\usepackage[latin1]{inputenc}
\usepackage{float}
\usepackage{color}
\usepackage{graphicx}
\usepackage{amssymb}
\usepackage{babel}
\usepackage{amsmath}
\usepackage{fancyhdr}
\usepackage{geometry}
\usepackage{comment}
\usepackage[normalem]{ulem}
\usepackage{mathtools}   
\usepackage{xcolor}
\usepackage{url}

\usepackage{dsfont}
\usepackage{braket}

\usepackage[affil-it]{authblk}

\author[1]{Angus Crookes}
\author[1]{Ben Yuen}
\author[2,3]{Stephen M. Hanham}
\author[1]{Angela Demetriadou\thanks{a.demetriadou@bham.ac.uk}}
\affil[1]{School of Physics and Astronomy, University of Birmingham, Edgbaston, Birmingham, B15 2TT, United Kingdom}
\affil[2]{Department of Materials, Imperial College London, South Kensington, SW7 2AZ, United Kingdom}
\affil[3]{School of Engineering, University of Birmingham, Edgbaston, Birmingham, B15 2TT, United Kingdom}

\title{
Multi-partite entanglement in extreme nanophotonic cavities  
}

\begin{document}

\maketitle


\begin{abstract}
	
Multi-partite entanglement is fundamental to emerging quantum technologies such as quantum networks, which ultimately require devices with strong light-matter interactions and long coherence times. Here, we introduce nanobeam photonic crystal cavities combining both extreme quality factors ($\sim10^{7}$) with sub-wavelength field confinement to reach unprecedented light-matter interactions.
Operating at $780$ nm, our devices are tailored for efficient coupling and entanglement with ultracold $^{87}$Rb atoms, a key ingredient in quantum networks due to their hyperfine structure. 
Our new designs also facilitate the precise optical trapping of atoms, and we demonstrate coherent entanglement generation between them, that is remarkably resilient to atomic displacements.
These platforms can be easily scaled-up to extremely large quantum networks, for distributed quantum computing and future light-based quantum technologies.

\end{abstract}

\section*{Introduction}

Quantum entanglement is fundamental to quantum computing and other emerging quantum technologies \cite{nielsen2002quantum, grynberg2010introduction, horodecki2009quantum}. In particular, the realisation of quantum networks, that exploit both local and global entanglement across separate quantum processors, has opened up powerful applications in distributed computing \cite{jiang2007distributed, wehner2018quantum}, sensing \cite{proctor2018multiparameter} and encryption \cite{pirandola2020advances}. 
Quantum networks consist of local nodes, used to manipulate and store quantum information, and quantum channels which connect the nodes together and distribute entanglement across the network \cite{kimble2008quantum, van2004quantum, reiserer2022colloquium}. In practice, photons (flying qubits) are often used to connect separate nodes due to their low absorption losses over large distances \cite{van2014quantum, duan2010colloquium, lines1984search}. Meanwhile, quantum emitters (stationary qubits) - such as trapped ions and cold atoms - are used to receive information and store it for long periods of time, due to the narrow linewidth of their hyperfine transitions \cite{duan2010colloquium, reiserer2015cavity, steck2001rubidium}. 
The efficiency of entanglement generation in a quantum network can be enhanced by increasing the coupling between flying and stationary qubits, which is essential for high-rate entanglement schemes based on quantum repeaters \cite{borregaard2019quantum, briegel1998quantum, muralidharan2016optimal}. This is commonly achieved using optical cavities, that increase light-matter interactions ~\cite{reiserer2022colloquium, reiserer2015cavity, lodahl2015interfacing, gonzalez2024light} through both spatial and temporal confinement of light. However, there is significant potential for further enhancement, especially in the visible regime, where cold atoms such as $^{87}$Rb, have their hyperfine transitions.

Indeed, many photonic devices that couple to cold atoms or other quantum emitters (such as quantum dots, fluorescent molecules etc.) have been realised - including micro-pillar \cite{reithmaier2004strong}, micro-disk \cite{peter2005exciton} and plasmonic nanocavities~\cite{chikkaraddy2016single}. In particular, nanobeam cavities have a unique design that also facilitates scaling up to extremely large quantum networks \cite{ohta2011strong,yoshie2004vacuum,mccutcheon2008design,thon2009strong}. However, achieving \textit{simultaneously} low losses (high quality factor) and sub-wavelength field confinement (small mode volume) is very challenging, due to material and scattering losses. 
Therefore, most optical cavities exist on the fringe of the strong coupling regime, and require precise positioning of quantum emitters using complex trapping mechanisms, to ensure efficient entanglement generation.
While some silicon nanobeam cavities do posses both sub-wavelength mode volumes and high quality factors \cite{hu2016design, hu2018experimental, ouyang2024singular}, they are challenging to implement in the optical regime to realize entanglement and quantum networks with cold atoms. Hence, the need for new designs operating efficiently in this region.

In this letter, we propose Si$_3$N$_4$ nanobeam photonic crystal cavities exhibiting \textit{simultaneously} unprecedented optical confinement ($Q = 10^7$) and extreme field enhancements ($V < 0.7 \lambda^3$) at $780$ nm to strongly couple and entangle $^{87}$Rb atoms.  In these devices, we demonstrate local multi-partite entanglement that is robust to fluctuations and displacements of atoms, while the nanobeam architecture is also scalable up to a large number of atoms, making them ideal for constructing quantum networks for both local and remote entanglement.

\section*{Extreme nanobeam cavities}
\begin{figure}
	\centering
	\includegraphics[width=1\textwidth]{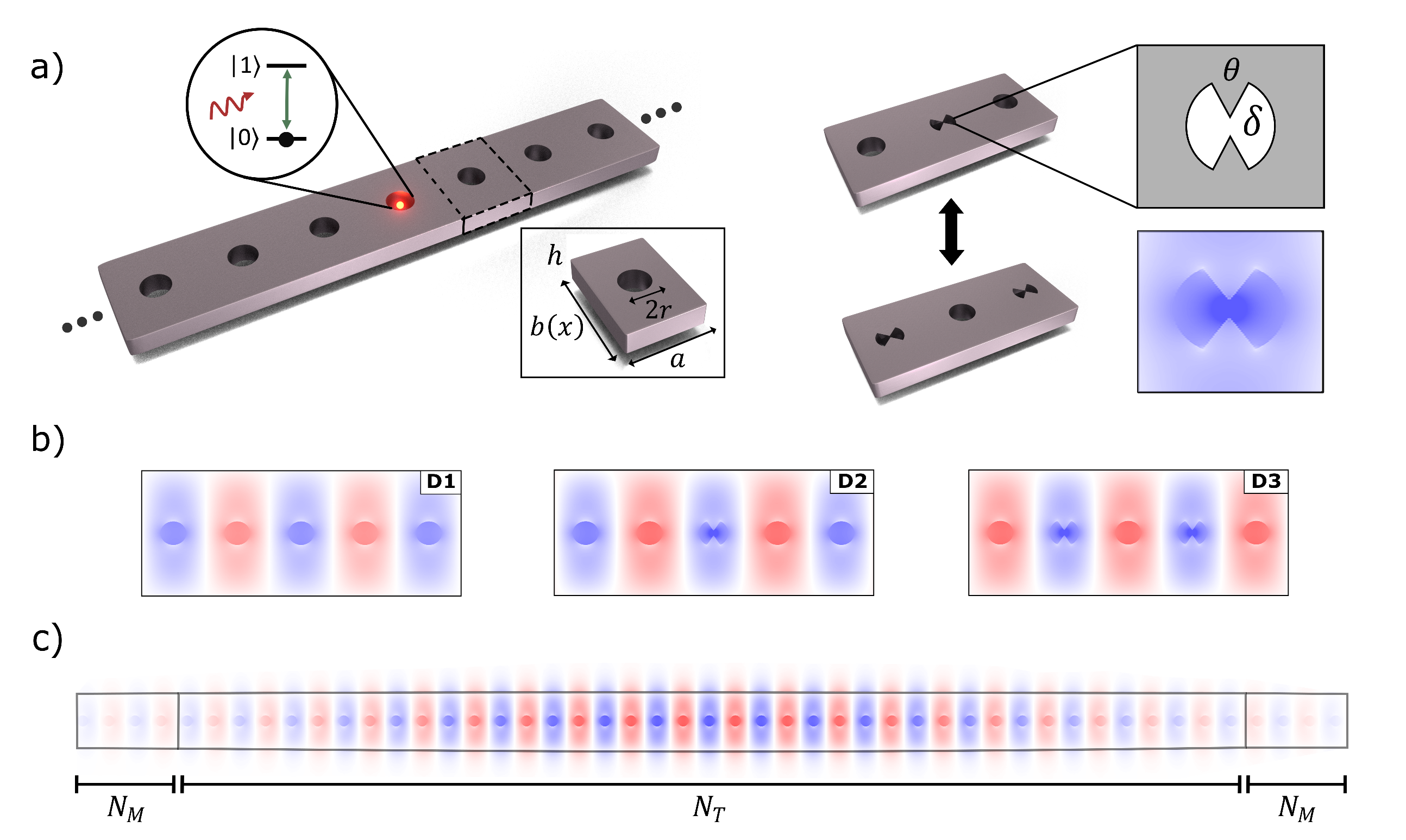}
	\caption{Schematic of nanobeam photonic crystal cavity designs. (a) Si$_3$N$_4$ nanobeam with lattice constant $a=261.87$ nm, height $h=300$ nm, spatially tapered width $b(x)$, and air hole radius $R=52.66$ nm. The dielectric tips are parameterised by their size $\theta=30^{\circ}$, tip-tip spacing $\delta=20$ nm, and rescaled hole radius $r_{\text{tip}} = 59.6$ nm. (b) Photonic mode profiles of all designs D1$\rightarrow$D3 operating at $\lambda=780$ . D1 contains no tips, while D2 and D3 contain one and two tips, respectively. (c) Profile of the resonant modes electric field (not to scale) for D1 with tapered region $N_T$ and mirror region $N_M$ showing confinement in the centre. }\label{fig:photonic_crystal}
\end{figure}

Here, we consider a nanobeam cavity made of Si$_3$N$_4$ with relative dielectric permittivity $\epsilon = 3.9(1 + i10^{-7})$~\cite{sinclair20201}, since it has a relatively high refractive index with small material losses. We have chosen to use a much higher loss tangent than reported in the literature~\cite{beliaev2022optical, vogt2016development}, to account for fabrication imperfections and surface roughness. Note that various optical properties for Si$_3$N$_4$  have been reported, since they strongly depend on the fabrication method used.
The waveguide height is $h=300$ nm, and patterned with periodic air holes of constant radius $r=53$ nm and lattice spacing $a=262$ nm.
The width $b(x)$ is tapered from the centre of the waveguide towards both ends (over 35 unit cells) and varies from $b_0 = 557$ nm to $b_{\text{end}} = 458$ nm with the nanobeam design (D1) shown in Figure \ref{fig:photonic_crystal}(a). In this way, the central region of the waveguide supports a $780$ nm photonic mode, while both nanobeam ends support a photonic band-gap that act as mirrors, trapping resonant photons and creating a high-finesse cavity (see Supplementary Information). 
The mode is further confined for larger mirror regions. In fact, for twenty mirror cells of width $b_{\text{end}}$ the radiative losses are negligibly small, and therefore cavity losses arise almost entirely from material absorption (see Supplementary Information). 

This nanobeam cavity (D1) mode at $\lambda = 780$ nm is $y$-polarized, with a schematic of the field distribution shown in Figure \ref{fig:photonic_crystal} (c). In addition, the mode concentrates energy in the air rather than the dielectric - well-suited for coupling with atoms optically trapped in each hole. The cavity retains photons for extremely long times, having an ultra-high quality factor $Q = 1.3 \times 10^7$, diffraction limited mode volume $V = 2.2~(\lambda/n)^3$, and extremely high cooperativity (assuming coupling to $^{87}$Rb atoms) $C=4.5\times 10^5$ where the mode volume is calculated using $V=\frac{1}{2}\iiint\left(\mathbf{D}\cdot\mathbf{E} + \mathbf{H}\cdot \mathbf{B}\right) dV ~/~ \frac{1}{2}\text{max}\left(\mathbf{D}\cdot{\mathbf{E}}\right)$ using the electric and magnetic fields $\mathbf{E}$ and $\mathbf{H}$ respectively. 
The mirror region of the nanobeam is already optimized to maximize the quality factor. 
Therefore, to further enhance the light matter interaction and cooperativity, we instead need to reduce the mode volume to sub-wavelength values. This is not easily achieved in optical cavities, since the electromagnetic response of photonic crystals is based on diffraction, and therefore are bound by the diffraction limit. This limits most optical cavities to exist on the fringe of the strong coupling regime, when coupled with cold atoms.

Here we produce two new designs with both extreme quality factors and mode volumes, by utilizing the evanescent fields at a high-index dielectric-air interface. 
The two designs use bow-tie tips in the central (D2) and off-central (D3) unit cells (see Figure 1(a),(b)) to create an evanescent field in which extreme sub-wavelength localisation is achieved at optical frequencies
. We find that two main parameters control the degree of field enhancement at the tips: (i) the size of the tip ($\theta$) and (ii) the tip-to-tip spacing ($\delta$).  In particular, wider tips (larger $\theta$) with a smaller separation (smaller $\delta$) produce greater field enhancements due to an increased overlap between the evanescent fields from each tip. However, to account for experimental constraints we instead choose a large tip size of $\theta=30^{\text{o}}$ and a tip-to-tip spacing of $\delta = 20$ nm for all our devices, which are comfortably within current nano-fabrication capabilities~\cite{ouyang2024singular}.

By adding the dielectric tip inside the unit cell hole, the photonic band shifts away from $780$ nm and therefore, introduces scattering losses. To tune the photonic band back to $780$ nm, one needs to re-scale the radius of the hole where bow-tie tips are present from $r_0=53$ nm $\rightarrow r_{\text{tip}}=60$ nm (see Supplementary Information). After this, radiative losses are again negligibly small and these two new designs retain their ultra-high quality factors of $Q=1.2\times10^7$ and $1.1\times10^7$ for design D2 and D3 respectively. The small reductions in the quality factor are due to unavoidable scattering losses introduced by the dielectric tips. The mode volume though, now takes sub-wavelength values $V=0.66~(\lambda/n)^3$ for both designs. Further reduction of the tip-to-tip spacing reduces the mode volume by at least one order of magnitude \cite{ouyang2024singular} (see Supplementary Information), but here we focus on immediately realizable designs with current fabrication techniques. The field distribution of the resonant mode of all three cavities are shown in Figure \ref{fig:photonic_crystal} (b) where the enhancement of the electric field ($E_{\text{y}}$) at the tips is clearly illustrated. This unique combination of high quality factor and sub-wavelength mode volume leads to unprecedented and extreme values for the cooperativity i.e. $C=1.3\times10^6$ and $C=1.2\times10^6$ for each cavity, making them ideal for efficient coupling and entanglement with quantum emitters, such as cold atoms and quantum dots. These new nanobeam designs are deep within the strong coupling regime for visible wavelengths, unlike current state-of-the-art photonic cavities that tend to operate at the boundary between weak and strong coupling.

\begin{figure}
	\centering
	\includegraphics[width=1\textwidth]{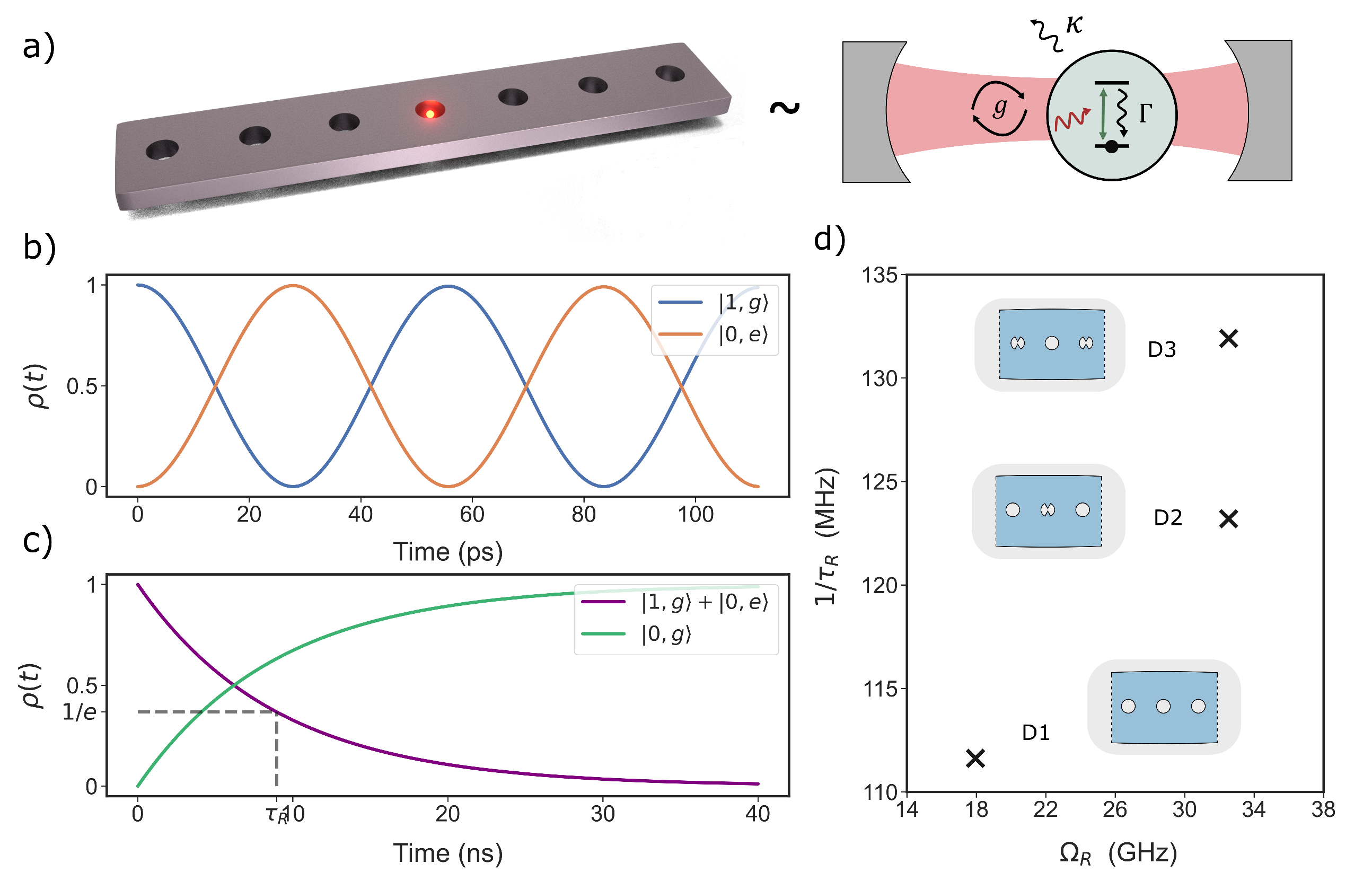}
	\caption{Schematic and dynamics for a single two-level Rb atom suspended in our nanobeam photonic crystal cavity designs (a) Representation of an atom in nanobeam cavity D1. The atom is coupled to a resonant photonic mode with coupling strength $g$ and with cavity and atomic loss factors $\kappa$ and $\gamma$. (b) Short (over $100$ ps) time-dynamics of a single QE interacting with the resonant photonic mode of the nanobeam cavity; unitary Rabi oscillations can be seen, occurring with frequency $g/\pi$. (c) Long (over $40$ ns) time-dynamics and decay lifetime, $\tau_R$, for the amplitude envelope of Rabi cycles. (d) Rabi frequency and decay lifetime comparison for a single quantum emitter placed in all three nanobeam designs.}
	\label{fig:singlequbit_dynamics}
\end{figure}

To maximize the light-matter interaction between the cavity and a quantum emitter, we optically trap $^{87}$Rb atoms \textit{inside} the central unit cell of each design. $^{87}$Rb atoms have their hyperfine transition $5^2S_{1/2}\rightarrow 5^2P_{3/2}$ at $780$ nm, for which we designed the photonic systems to operate. 
An $^{87}$Rb atom is trapped in the central hole, or between the bow-tie tips of each design, using a linear combination of blue-detuned modes $U_{\text{total}}(\mathbf{r}) = U_1(\mathbf{r}) + U_2(\mathbf{r})$ where $U_1$ and $U_2$ create a repulsive trap by repelling the atoms from regions of high electric field intensity along the $x$ and $y$ directions respectively (see Supplementary Information for more details). This creates a strong quadratic trap with average displacements $\Delta x^{(1)}_0=5.3$ nm, and $\Delta y^{(1)}_0 = 5.1$ nm for nanobeam cavity D1, and $\Delta x^{(2)}_0=4.8$ nm, and $\Delta y^{(2)}_0 = 7.5$ nm for nanobeam D2.

To demonstrate the effect of such strong light-matter interactions in these systems with cold atoms, we use an open quantum system formalism.
The time evolution is governed by a Lindblad Master Equation \cite{breuer2002theory, manzano2020short} where the density operator $\rho$ evolves under:
\begin{equation}
	\frac{d\rho}{dt} = -i\left[\mathcal{H},\rho\right] + \kappa\left(a\rho a^{\dag} - \frac{1}{2}\left\{a a^{\dag},\rho\right\}\right) + \gamma\sum_{i=1}^{N}\left(\sigma_i\rho \sigma_i^{\dag} - \frac{1}{2}\left\{\sigma_i \sigma_i^{\dag},\rho\right\}\right)
	\label{lindbladme}    
\end{equation}
with $N$ atom   Tavis-Cummings Hamiltonian \cite{shore1993jaynes, dicke1954coherence, garraway2011dicke}:
\begin{equation}
	\mathcal{H} = \hbar\omega_ca^\dag a + \frac{\hbar\omega_0}{2}\sum_{i=1}^{N}\sigma_i^z + \hbar \sum_{i=1}^Ng_i(\mathbf{r}_i)\left(a\sigma_i^{\dag} + a^\dag\sigma_i\right)
	\label{dicke-model-ham}
\end{equation}
where $a^{\dag}$ and $a$ are the creation and annihilation operators for the photonic mode with frequency $\omega$ and dissipation rate $\kappa$, and $\sigma^{\dag}$ and $\sigma$ are the atomic raising and lowering operators for a Rb atom with frequency $\omega_0$ and linewidth $\gamma$ \cite{steck2001rubidium}. Importantly, the coupling strength between the photonic mode and each atom is parameterised by $g(\mathbf{r}_i) = |\boldsymbol{\mu}|\sqrt{\frac{\omega_c}{\hbar\epsilon_0\epsilon V(\mathbf{r}_i)}}$ where $\boldsymbol{\mu}$ is the atomic dipole moment, $\epsilon_0$ the permittivity of free space, and $V(\mathbf{r}_i)=\frac{1}{2}\iiint\left(\mathbf{D}\cdot\mathbf{E} + \mathbf{H}\cdot \mathbf{B}\right) dV ~/~ \frac{1}{2}\mathbf{D}(\mathbf{r}_i)\cdot{\mathbf{E}}(\mathbf{r}_i)$ is the corresponding (position dependent) mode volume. 

To demonstrate the extreme coupling in our devices, we first consider a single Rb atom trapped within the central unit cell of nanobeam cavity D1 i.e. trapped at $\mathbf{r}_1=(0,0,0)$ and initialised in a state with a single photon in the cavity as shown in Figure \ref{fig:singlequbit_dynamics} (a). Figure \ref{fig:singlequbit_dynamics} (b) and (c) illustrate the corresponding quantum dynamics for two different time-scales i.e. the populations of $|1,g\rangle$ and $|0,e\rangle$ over $100$ ps and $40$ ns, respectively. Rabi oscillations emerge with frequency $\Omega_R = g/\pi$ as photons are able to successfully complete hundreds of cyclic energy exchanges between the atom and the cavity, before being significantly impacted by any losses. 
Indeed, the ultra-high $Q$-factor and large interaction strength pushes this device comfortably into the strong coupling regime i.e. $g = 9$ GHz which is approximately three orders of magnitude larger than the cavity loss i.e. $\kappa = 29$ MHz. Therefore, the coherent oscillations are preserved for a long time with lifetime of $\tau_{\text{R}} \approx 10$ ns, which is  much longer than most current cavity designs at visible wavelengths.

Although nanobeam design D1 already achieves strong coupling, one may need to push deeper into the strong coupling regime for more rapid interactions and entanglement. The new nanobeam designs D2 and D3 make use of bow-tie tips to increase the coupling strength, since the mode volume has been reduced to subwavelength values. In this system, the Rb atom is optically trapped between the two tips, again using a linear combination of blue-detuned modes (see Supplementary Information).
Figure  \ref{fig:singlequbit_dynamics} (d) shows the frequency of Rabi oscillations is almost doubled in both designs, while the lifetime is not significantly affected. Hence, both nanobeam designs D2 and D3 operate much further into the strong coupling regime than D1, which was so far inaccessible. Note that one can further reduce the bow-tie tip spacing to achieve even higher coupling strengths. Hence, light-matter interactions are enhanced to unprecedented levels in the optical regime.

\section*{Multi-partite entanglement}

\begin{figure}
\centering
\includegraphics[width=1\textwidth]{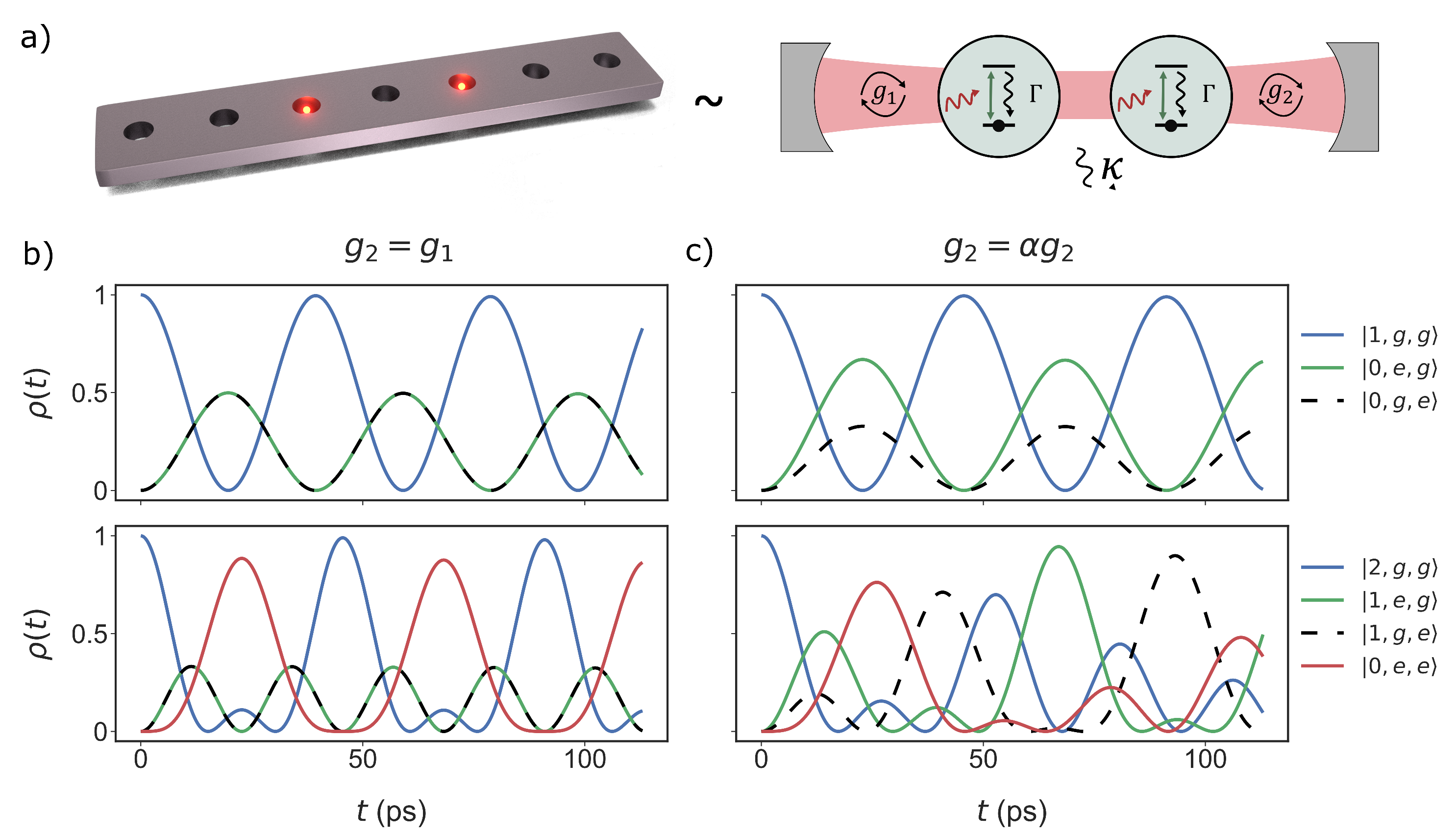}
\caption{Schematic and quantum dynamics for two $^{87}$Rb atoms interacting with the resonant nanobeam mode. (a) Representation of multiple atoms in the nanobeam cavity D1, coupled to the resonant mode with strengths $g_1$ and $g_2$ and with cavity and atomic loss factors $\kappa$, $\gamma$ respectively. The quantum dynamics are shown for equal i.e. $g_1=g_2$ (left) and unequal i.e. $g_2=\alpha g_1$ (right) coupling of both quantum emitters to the photonic mode, when the system is initially prepared in the (b, c) one photon state $\ket{100}$, and (d, e) two photon state $\ket{200}$. The coupling parameter $\alpha=0.7$ is chosen to demonstrate a situation where entanglement is significantly reduced. }
\label{fig:multiqubit_dynamics}
\end{figure}


We now use this extreme light-matter interaction to realise multi-partite entanglement between two Rb atoms within the new nanobeam cavities. Initially, we consider two Rb atoms collectively interacting with the photonic mode of design D1, as shown in Figure \ref{fig:multiqubit_dynamics} (a). In particular, we focus on atoms at positions $\mathbf{r}_{1,2} = (\pm a,0,0)$ such that the coupling of both atoms to the photonic mode is identical i.e. $g_1(\mathbf{r}_{1})=g_2(\mathbf{r}_2)$ as this serves a valuable benchmark for subsequent arrangements. The generation of periodic entanglement is shown in Figure ~\ref{fig:multiqubit_dynamics} (b). Here, there is a coherent exchange of energy between the separable state $\ket{\chi_0^{(1)}} = \ket{1}\ket{gg}$ and the maximally entangled atomic state $\ket{\chi_1^{(1)}}=\frac{1}{\sqrt{2}}\ket{0}\otimes\left(\ket{eg}+\ket{ge}\right)$ due to the high quality factor of our cavity. Importantly, the strong coupling dynamics are also more pronounced than with a single atom, such that $\ket{\chi_1^{(1)}}$ is generated periodically at frequency $\sqrt{2}\Omega_R$ where $\sqrt{2}$ is the Dicke enhancement. 

\begin{figure}
\centering
\includegraphics[width=1\textwidth]{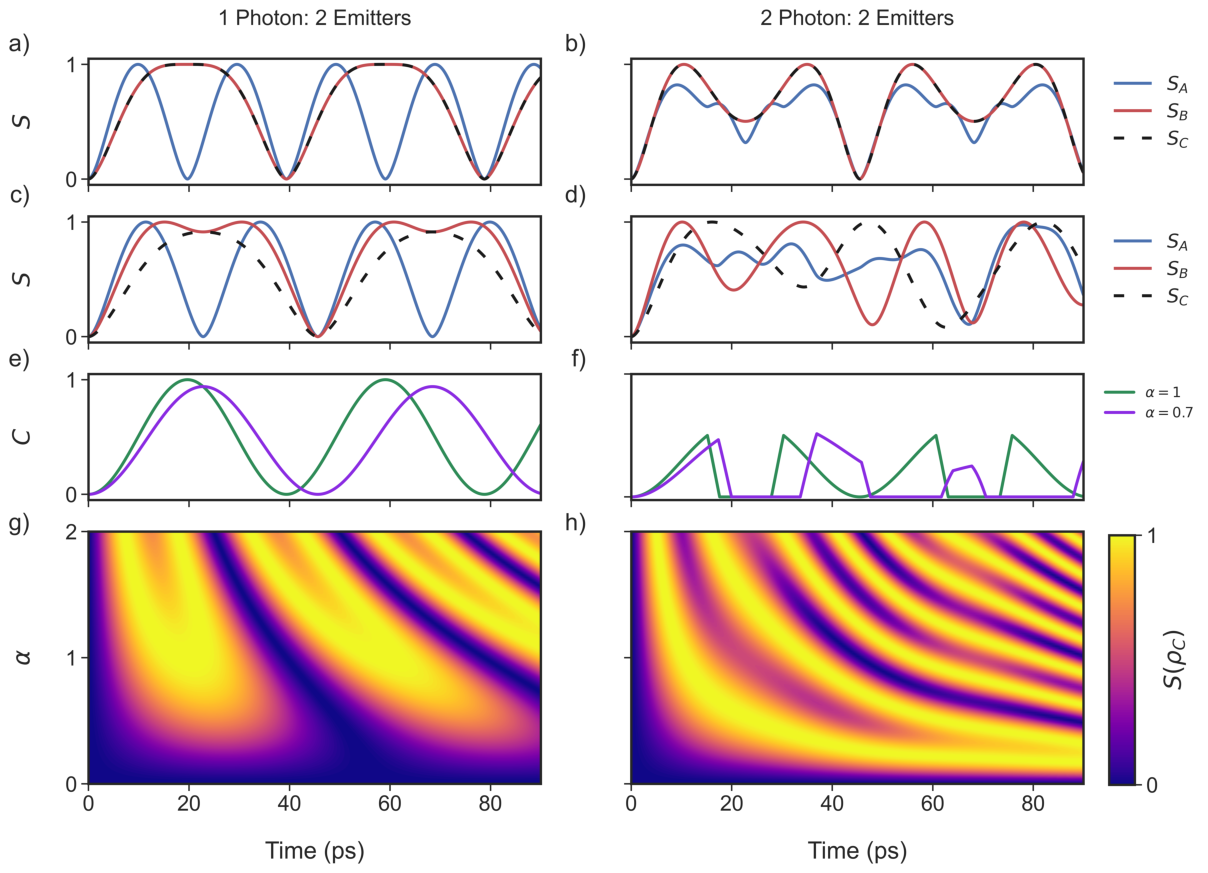}
\caption{Quantum correlations between two quantum emitters interacting with the photonic mode of nanobeam cavity D1. The system is initialised in the one photon (left) and two photon (right) initial states.  (a, b) Entropy ($S$) for equal coupling $g_2=g_1$ (c, d) Entropy ($S$) for unequal coupling $g_2=\alpha g_1$ where $\alpha=0.7$. The entropy, $S_i$, is calculated for all three subsystems $i\in\{A,B,C\}$ where $A,B,C$ are the cavity and atomic subsystems respectively (e, f) Concurrence, $C$, between both emitters. (g, h) Dependence of  $S_C$, on the coupling asymmetry $\alpha$.}
\label{fig:quantumcorrelations}
\end{figure}

Although the quantum dynamics shown in Figure \ref{fig:multiqubit_dynamics} (b) demonstrate the generation of entangled states, the von Neumann entropy and the concurrence are better measures of the entanglement \cite{nielsen2002quantum, wootters2001entanglement}. The von Neumann entropy characterises entanglement between any two partitions of the system, and is expressed as $S_i(\rho_i) = - \text{Tr}\left(\rho_i\log\rho_i\right)/\ln(d)$ for the reduced density matrix  $\rho_i$ of subsystem $i\in\{A,B,C\}$ with dimension $d$, where $\rho_{A}$ is the photonic subspace and $\rho_{B}$ and $\rho_{C}$ are the two atomic subspaces respectively \cite{nielsen2002quantum}. The concurrence measures entanglement directly between the atomic subsystems, and is expressed as $C(\rho) = \text{max}\left(0, \lambda_1-\lambda_2-\lambda_3-\lambda_4\right)$ where $\{\lambda_i\}$ is the set of ordered eigenvalues of $R = \sqrt{\sqrt{\rho}\tilde{\rho}\sqrt{\rho}}$ and $\tilde{\rho} = \sigma_y\otimes\sigma_y\rho^*\sigma_y\otimes\sigma_y$ is the spin flipped density matrix \cite{wootters2001entanglement}. Importantly, both measures are unity for maximally entangled subspaces, and zero for completely separable subspaces. 
The entropy and concurrence for two atoms interacting with the resonant mode of nanobeam design D1 is shown in Figure \ref{fig:quantumcorrelations} (a),(e) for a single photon excitation. The maximum atomic entanglement is obtained when $\ket{\chi_1^{(1)}}$ is fully populated (i.e. when $S_A=0$ and both $S_B=S_C=1$). Note that the photon is separable from the atoms when the system is entirely in the state $\ket{\chi_0^{(1)}}$ or $\ket{\chi_1^{(1)}}$ - which is why the oscillations in $S_A$ are twice as fast.

Importantly, the coupling strength of each atom to the photonic mode is dependent on each atom's position within the air hole, and can vary depending on the properties and depth of the optical trap keeping the atoms in place. Therefore, to account for imperfect and unpredictable variations in the optical traps, as well as possible atomic movements, we also consider the case where the system loses its translational invariance i.e. $g_1(\mathbf{r}_1) \neq g_2(\mathbf{r}_2)$.  To demonstrate the general effect this has on entanglement, we first assume one atom interacts with the photonic mode with strength $g_1=g(\mathbf{r}_1)$ while the second has strength $g_2 = \alpha g_1$, where $\alpha\in \mathbb{R}$ is an arbitrary constant. The quantum dynamics when $\alpha = 0.7$ are shown in Figure \ref{fig:multiqubit_dynamics} (c) with $\alpha$ chosen to illustrate a situation where entanglement is significantly reduced. Note however, that within both nanobeam designs (D1 and D3) the maximum difference in coupling strength is defined by the trap depth (see Supplementary Information) which gives $\alpha=0.99$ and $\alpha=0.96$ respectively. Figure \ref{fig:multiqubit_dynamics} (c) shows there are now coherent oscillations between the initial state $\ket{\chi_0^{(1)}} = \ket{1}\ket{gg}$ and the entangled state $\ket{\chi_1^{(1)}}=\ket{0}\otimes\frac{1}{\sqrt{g_1^2 + g_2^2}}\left(g_1\ket{eg} + g_2\ket{ge}\right)$ at frequency $\sqrt{g_1^2 + g_2^2}\Omega_R$ (see Supplementary details for calculations). Notably, the unequal coupling produces a splitting between the different excitation probabilities of each atom (despite remaining in phase) with splitting magnitude $\Delta_s = |1-\alpha^2|/|1+\alpha^2|$. In addition, the entanglement of $|\chi_1^{(1)}\rangle$ is reduced from the maximally entangled state $|\psi_+\rangle = \ket{0}\otimes \frac{1}{\sqrt{2}}\left(\ket{eg}+\ket{ge}\right)$ according to the entanglement fidelity $\langle \psi_+|\chi_1^{(1)}\rangle = (1+\alpha) / \sqrt{2(1+\alpha^2)}$  which has maximum fidelity only when the coupling is identical for the two emitters i.e. $\alpha=1$.

The von Neumann entropy and concurrence of this system with unequal coupling ($\alpha=0.7$) is shown in Figure \ref{fig:quantumcorrelations} (c),(e) for a single photon excitation. Here, the atomic entangled state is formed when $S_A=0$ and $S_B, S_C > 0$, but the degree of entanglement is reduced due to the larger population of atom one.
For $\alpha>1$ the second atom experiences stronger coupling as shown in Figure. \ref{fig:quantumcorrelations} (g) and also the rate of entanglement formation increases due to a larger value of $\sqrt{g_1^2+g_2^2}$.
For extreme variations in the coupling strength, we obtain single atom strong coupling where $S_C\rightarrow 0$ as $\alpha \rightarrow 0$ and correspondingly $S_B\rightarrow 0$ when $\alpha >> 1$. 
Finally, note that an entangled state is also generated when $N$ atoms are coupled to the resonant mode. In this situation, the multi-partite entangled state can be expressed as $|\chi_1^{(1)}\rangle = \ket{0}\otimes\frac{1}{\sqrt{\sum g_i^2}}\sum_{i=1}^N g_i\sigma_i^{\dag}|g\rangle^{\otimes N}$ and has time evolution $\propto \sin^2\left(t\sqrt{\sum_{i=1}^N g_i^2}\right)$. In parallel to the two atom case, $|\chi_1^{(1)}\rangle$ becomes the maximally entangled $|W\rangle$ state when the coupling of all $N$ Rb atoms to the photonic mode is equivalent (see Supplementary Information).

To gain deeper insights into the impact of unequal coupling, we examine the system when it is initialized in the two-photon state $\ket{\chi_0^{(2)}} = \ket{2}\ket{gg}$. In general, energy exchange now occurs between a set of four orthogonal states: $\ket{\chi_0^{(2)}}$, $\ket{\chi_1^{(2)}}=\ket{1}\otimes\frac{1}{\sqrt{g_1^2 + g_2^2}}\left(g_1\ket{eg} + g_2\ket{ge}\right)$,  $\ket{\chi_2^{(2)}}=\ket{0}\ket{gg}$ and $\ket{\chi_3^{(2)}} = \ket{1}\otimes\frac{1}{\sqrt{g_1^2 + g_2^2}}\left(g_2\ket{eg} - g_1\ket{ge}\right)$. When $\alpha=1$, $\ket{\chi_3^{(2)}}$ decouples from the system, producing oscillations between the remaining triplet states \cite{garraway2011dicke}. Figure \ref{fig:multiqubit_dynamics} (d) shows the bare state population dynamics when the coupling is equal i.e. $\alpha=1$. Here, the $\ket{1eg}$ and $\ket{1ge}$ states oscillate in phase and with equal amplitude, as they exist only as a part of $\ket{\chi_1^{(2)}}$. 
Figure \ref{fig:quantumcorrelations} (b) shows the corresponding quantum correlations between all three subsystems when $\alpha = 1$. In this case, the entropies all oscillate at the same frequency, and are only separable in the state $|\chi_0^{(1)}\rangle=|2gg\rangle$. In addition, while the entropy of both atomic subsystems can be maximised, this does not correspond to maximum atomic entanglement as there are still correlations with the photon i.e. $0 < S_A < 1 $. This is shown clearly in Figure \ref{fig:quantumcorrelations} (f) where $C<1$ and features periodic regions of complete disentanglement, during which the remaining entanglement is entirely with the photonic mode.
Conversely, Figure \ref{fig:multiqubit_dynamics} (e) shows the evolution when $\alpha=0.7$, where due to coupling with $\ket{\chi_3^{(2)}}$ the bare state populations now differ in amplitude and oscillation frequency and no periodic entanglement is generated. Furthermore, the correlations between each subsystem  are also sensitive to unequal coupling as shown in Figure \ref{fig:quantumcorrelations} (d), (f), (h). Indeed, entanglement entirely between both atoms is much harder to consistently form in systems with higher excitation numbers. For this reason, a single photon excitation is much more advantageous for realising entanglement within nanophotonic cavities.
\newline

\begin{figure}
	\centering
	\includegraphics[width=1\textwidth]{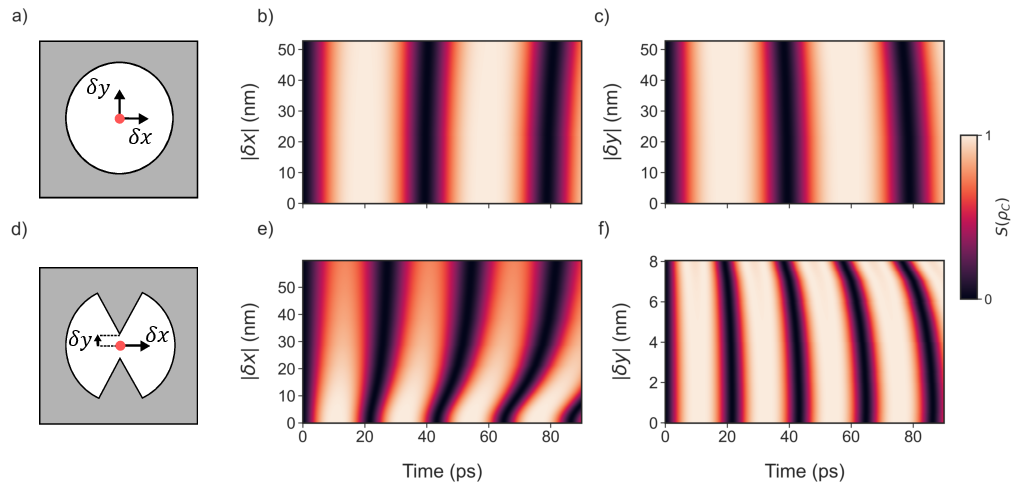}
	\caption{ \baselineskip 13pt   Characterisation of entanglement between two atoms interacting with the resonant mode of each nanobeam photonic crystal cavity design at positions $\mathbf{r}_1=(-a,0,0)$ and $\mathbf{r}_2=\left(a+\delta x,\delta y,0\right)$ respectively. Entropy ($S_C$) for atom two as a function of position, $|\delta x|$ and $|\delta y|$, inside (a,b,c) D1 and (d,e,f) D3.}
	\label{specificdesigns}
\end{figure}

Our results so far demonstrate that maximum entanglement between both atoms occurs when $g_1(\mathbf{r}_1)=g_2(\mathbf{r}_2)$ - or when values are as close to this as possible. For our new designs, the atoms are optically trapped at the centre of each air hole, with small spatial variations that give rise to differences in their coupling strengths, see Figure \ref{specificdesigns} (a),(d). Therefore, to exhaustively characterise the entanglement within our nanobeam designs, we investigate the spatial variation of the coupling strength and its impact on quantum correlations within nanobeam cavities D1 and D3.
In particular, we fix the position of the first atom such that $\mathbf{r}_1 = (-a,0,0)$ and displace the second by $\delta x$ and $\delta y$ such that $\mathbf{r}_2 = (a+\delta x,\delta y, 0)$ as shown in Figure \ref{specificdesigns} (a,b,c) and \ref{specificdesigns} (d,e,f) for D1 and D3 respectively. In nanobeam cavity D1 the displacement creates a variation in the coupling strength up to $\alpha = 0.95$ and $\alpha = 1.06$ for $\delta x = r$ and $\delta y = r$ respectively. Therefore, attributed to the minute change in coupling strength, the entanglement in this system is stable - decreasing by only $\sim 0.2\%$ in the extreme and very unlikely case when displacement of one atom reaches the dielectric interface, while the other remains at the air hole centre.
For nanobeam cavity D3 the field is more localised at the dielectric tips, which produces larger differences between the two coupling strengths up to $\alpha = 0.52$ and $\alpha = 0.8$ for $\delta x=r$ and $\delta y=\delta$ respectively.
However, although there is a strong variation in $g$ along the unit cell, the optical trap confines the atom within a small region where entanglement decreases by no more than  $2.7\%$  (see Supplementary Information for details on the optical trap). This suggests that both nanobeam designs can create periodic, local entanglement, and provide a realistic, robust platform for entanglement in quantum networks \cite{kimble2008quantum, cirac1997quantum}. In the future, we envision similar designs using these nodes, connected together to scale up to a massive network, with external pump fields controlling the state of the atoms and transmission of photons between each node achieved through tailoring the mirror region size along a specified direction. In such systems, local storage of quantum information is performed by atoms, and long distance transmission between nodes achieved via photons coupled out of the cavity via the waveguide-like design of the nanobeam system.

In conclusion, we have presented the design and characterisation of nanobeam photonic crystal cavities commanding unprecedented optical confinement and strong field enhancements, $Q = 10^7$ and $V < 0.7 \lambda^3$, while operating at $780$ nm for strong coupling with cold atoms such as $^{87}$Rb. Unlike most other cavities, our designs exist deep into the strong coupling regime, where we demonstrate the generation of robust local multi-partite entanglement between the atoms. The stable, coherent and robust entanglement generated, in combination with the possibility to massively scale-up the photonic environment, is promising to realise extremely large-scale quantum networks.

\newpage

\bibliographystyle{unsrt}


\section*{Acknowledgements}

AD gratefully acknowledges support from the Royal Society University Research Fellowship URF\textbackslash R1\textbackslash 180097 and URF\textbackslash R\textbackslash 231024, Royal Society Research grants RGS \textbackslash R1\textbackslash 211093, funding from ESPRC grants  EP/Y008774/1 and EP/X012689/1.
AD, BY acknowledge support from Royal Society Research Fellows Enhancement Award RGF \textbackslash EA\textbackslash 181038, and AD, AC acknowledge funding from EPSRC for the CDT in Topological Design EP/S02297X/1.

\section*{Authors statement}

All the authors have accepted
responsibility for the entire content of this submitted
manuscript and approved submission.

\end{document}